\documentclass[twocolumn,showpacs,preprintnumbers,amsmath,amssymb]{revtex4}
\usepackage{graphicx}
\begin{document}
\def\be{\begin{equation}}
\def\ee{\end{equation}}
\def\bearr{\begin{eqnarray}}
\def\eearr{\end{eqnarray}}
\def\tc{$T_c~$}
\def\tcl{$T_c^{1*}~$}
\def\c2{ CuO$_2~$}
\def\ruo{ RuO$_2~$}
\def\lsco{LSCO~}
\def\bi{bI-2201~}
\def\tl{Tl-2201~}
\def\hg{Hg-1201~}
\def\sro{$\rm{Sr_2 Ru O_4}$~}
\def\rc{$RuSr_2Gd Cu_2 O_8$~}
\def\mgb{$MgB_2$~}
\def\pz{$p_z$~}
\def\ppi{$p\pi$~}
\def\sqo{$S(q,\omega)$~}
\def\tperp{$t_{\perp}$~}
\def\he4{${\rm {}^4He}$~}
\def\ags{${\rm Ag_5 Pb_2O_6}$~}
\def\nxcob{$\rm{Na_x CoO_2.yH_2O}$~}
\def\lsco{$\rm{La_{2-x}Sr_xCuO_4}$~}
\def\lco{$\rm{La_2CuO_4}$~}
\def\lbco{$\rm{La_{2-x}Ba_x CuO_4}$~}
\def\half{$\frac{1}{2}$~}
\def\thalf{$\frac{3}{2}$~}
\def\tst{${\rm T^*$~}}
\def\tch{${\rm T_{ch}$~}}
\def\jeff{${\rm J_{eff}$~}}
\def\nbc{${\rm LuNi_2B_2C}$~}
\def\cabc{${\rm CaB_2C_2}$~}
\def\nboo{${\rm NbO_2}$~}
\def\voo{${\rm VO_2}$~}
\def\nip{$\rm LaONiP$~}
\def\nisb{$\rm LaONiSb$~}
\def\nibi{$\rm LaONiBi$~}
\def\fep{$\rm LaOFeP$~}
\def\cop{$\rm LaOCoP$~}
\def\mnp{$\rm LaOMnP$~}
\def\fesb{$\rm LaOFeSb$~}
\def\febi{$\rm LaOFeBi$~}
\def\efeas{$\rm LaO_{1-x}F_xFeAs$~}
\def\hfeas{$\rm La_{1-x}Sr_xOFeAs$~}
\def\hSfeas{$\rm Sm_{1-x}Sr_xOFeAs$~}
\def\hCefeas{$\rm Ce_{1-x}Sr_xOFeAs$~}
\def\feas{$\rm LaOFeAs$~}
\def\Ndfeas{$\rm NdOFeAs$~}
\def\Smfeas{$\rm SmOFeAs$~}
\def\Prfeas{$\rm PrOFeAs$~}
\def\refeas{$\rm REOFeAs$~}
\def\refesb{$\rm REOFeSb$~}
\def\refebi{$\rm REOFeBi$~}
\def\ttog{$\rm t_{2g}$~}
\def\fese{$\rm FeSe$~}
\def\fete{$\rm FeTe$~}
\def\eg{$\rm e_{g}$~}
\def\dxy{$\rm d_{xy}$~}
\def\dzx{$\rm d_{zx}$~}
\def\dzy{$\rm d_{zy}$~}
\def\dxsq{$\rm d_{x^{2}-y^{2}}$~}
\def\dzsq{$\rm d_{z^{2}}$~}
\def\LAO{$\rm LaAlO_3$~}
\def\STO{$\rm SrTiO_3$~}
\def\h2s{$\rm H_2 S$~}
\def\smb{$\rm SmB_6$~}
\def\Smb{$\rm {\bf {SmB_6}}$}

\title{Majorana Fermi Sea in Insulating \Smb: \\
A proposal and a Theory of Quantum Oscillations in Kondo Insulators}
\author{ G. Baskaran}
\affiliation
{The Institute of Mathematical Sciences, C.I.T. Campus, Chennai 600 113, India\\
Perimeter Institute for Theoretical Physics, Waterloo, Ontario, Canada N2L 2Y5}

\begin{abstract}
In an exciting development, \smb, a Kondo insulator has been shown to exhibit bulk quantum oscillations.  We propose that \smb is a  \emph{bulk scalar Majorana Fermi Liquid} (MFL) with a finite gap for charge and spin excitations. In their study of Kondo insulators in 1993, Coleman, Miranda and Tsvelik (CMT) envisaged such a remarkable possibility, using a mean field ansatz. We generalize CMT theory to non-zero magnetic fields and show a counter intuitive result that the scalar Majorana fermi liquid, while remaining electrically insulating, responds to external magnetic field and exhibits Landau diamagnetism and quantum oscillations. Physics of an \emph{emergent compactified 2-channel Kondo lattice physics} that is behind formation of the novel scalar Majorana fermi liquid phase is discussed. It is also suggested that a known resistivity saturation in \smb as well as a new strong deviation of quantum oscillation amplitude from Lifshitz-Kosevich formula in \smb at low temperatures are due to a coherent fluctuation of charge of a neutral scalar Majorana fermion. Possible presence of 2-dimensional Majorana fermion excitations in surfaces of SmB6, and other Topological Kondo Insulators (TKI) is pointed out.
\end{abstract}

\maketitle
Correlated electron systems continue to surprise us and spring forth unexpected phenomena.  In this context, conceptually and physically simple experiments often reveal hidden truths about nature. Historically, low temperature specific heat measurements revealed to us the quantum secret of lattice vibrations. Same specific heat revealed fermi seas embedded in simple metals. In the same way magneto torque measurements, as dHvA quantum oscillations \cite{ShoenbergBook}, revealed to us shapes of fermi surfaces in k-space. They have been invaluable in understanding heavy fermions, Kondo insulators \cite{Hewson,AeppliFiskReviewKondoInsulators1992}, high Tc cuprates  and so on.

A recent experimental observation \cite{SuchitraSmB6QOsc}  of quantum oscillations in \smb, a Kondo insulator \cite{TedGeballe1969SmB6},  in magneto torque measurements has come as a surprise and challenge. Quantum oscillations are known to occur in metals that support charged quasi particles and fermi surfaces. \smb is an insulator, albeit with a small charge gap. This oscillation is seen at low temperatures, where resistivity is high indicating an electrically insulating ground state. Meanfree paths of quasi particles that are causing quantum oscillations,  as inferred from experiments, are anomalously large $\sim$ few microns.

As an explanation for the observed quantum oscillations, we propose that \smb supports a neutral scalar bulk Majorana fermi liquid with a fermi surface, making use of an early theory of Coleman, Miranda and Tsvelik \cite{ColemanMirandaTsvelik}. If proved correct this will be a first clean example of a bulk Majorana fermi sea in a condensed matter system.

A Majorana fermi sea \cite{TsvelikBook,ColemanMirandaTsvelik} has a zero energy (fermi) surface in k-space. However it has \emph{only particle like positive energy excitations}, which are complex canonical fermions. \emph{Hole like excitations or antiparticles are simply absent}. \emph{Such a one component free Majorana fermi sea can not be obtained by filling single particle states below a chemical potential}. Currently popular Majorana zero energy mode \cite{MajoranaZeroMode} (in the context of topological quantum computation) is a scalar fermion; pair of such zero modes forms a complex canonical fermion.

Our proposal becomes interesting in the background of an intense search for localized Majorana zero modes at, vortex cores, chain ends, surfaces, heterostructures \cite{MajoranaReview}. There has been search for Majorana Fermi sea \cite{SanthoshShankarGB} in Kitaev models and real systems (Kitaev-Khaliulin-Jackeli materials) \cite{KhaliulinJackeliIridates}  etc.  Isolating Majorana fermions or inducing zero Majorana modes for manipulation for topological quantum computation should be feasible at surfaces of \smb and suitably engineered heterostructures. Being a Topological Kondo Insulator \cite{ColemanTKI}, \smb is likely to have Majorana modes on its surfaces, edges and defects.

Qantum oscillations in a metal arises from magnetic field induced Landau quantization of one electron states. Kondo insulator, with a finite charge gap ($>> \hbar \omega_c$), is not expected to behave like a normal fermi liquid, in magneto torque measurements. Fermi surfaces seen in \smb strongly resemble those of isostructural and isovalent metallic counterparts LaB$_6$ and CeB$_6$. Temperature dependence of oscillation amplitudes are anomalous however and does not follow the Lifshitz Kosevich fromula. We call \smb as \emph{longitudinal insulator and transverse metal.} Another important known experimental result \cite{NeutronNMR} is presence of a small spin gap, in addition to charge gap. Optics in this material is also interesting and strongly temperature dependent \cite{optics}.

Using the above we make two straight forward inferences:  i) presence of low energy \emph{zero spin quasi particles that are charge neutral} and ii) presence of fermi surfaces, some what similar to spinful fermi liquid counterparts  LaB$_6$ and CeB$_6$. We envisage three possibilities. First one is that \smb is a quantum spin liquid state that has a spinon pseudo fermi surface. Spin gap excludes this possibility. Second possibility is presence of a gapless spin singlet excitations with a pseudo fermi surface and a gap for spin-1 excitations; this is suspected to be present in certain frustrated  spin half Heisenberg antiferromagnets.

A third possibility utilizes equivalence of a half filled band of spinful free fermions in a bipartite lattice with nearest neighbor hopping, to four Majorana fermion species \cite{ColemanMirandaTsvelik}, a scalar and a 3-vector, with identical zero energy (fermi) surfaces. In this scenario Kondo spins effectively interact with vector Majorana fermions and opens gap at the fermi level, \emph{leaving the scalar Majorana fermi sea unaffected}. That is, a spinless neutral Majorana fermi liquid (MFL) emerges from a parent fermi liquid, preserving the fermi surface features.

Rest of this article is about the remarkable possibility of survival of the scalar Majorana fermi sea of the conduction electron systems, inspite of interaction with Kondo spins. A seminal work \cite{ColemanMirandaTsvelik} of Coleman, Miranda and Tsvelik (CMT), that came in the wake of odd-frequency pairing \cite{BalatskyAbrahams}, indeed suggested this possibility. Their work uses an odd frequency meanfield pairing ansatz. In the present article we achieve this, without an explicit discussion about odd frequency pairing.

It was noticed by Coleman, Ioffe and Tsvelik \cite{CIT} that the odd pairing mean field theory of CMT has a close connection to a compactified 2-channel Kondo lattice problem. The two channels were identified to be the local spin and local (charge) pseudo spin degree of freedom of the single half filled conduction electron band. We offer a physical explanation for this emergent 2-channel Kondo physics in the present article.

What is not obvious in the work of CMT is the possibility of quantum oscillations arising from scalar neutral MFL. Infact, first reaction will be to say that such an electrically insulating liquid should not respond to uniform magnetic field . One of our central results is that the electrically insulating MFL indeed responds to uniform magnetic fields, in a way qualitatively similar to a charged fermi liquid, through changes in ground state energy. That is the \textbf{scalar Majorana fermion liquid exhibits Landau diamagnetism}.

\smb has a simple cubic structure, formed by B$_6$  octahedra bonded between vertices. Sm atoms occupy the body center. Each
$({\rm B_6)}^{2-}$ octahedron accepts two electrons and Sm$^{2+}$ ion fluctuates between two configurations 4f$^6$ 5d$^0$ and 4f$^5$ 5d$^1$.  A minimal model \cite{ColemanTKI,Kasuya} including spin-orbit coupling isolates one Kramers doublet each from 4f and 5d shells. They hybridize with each other, at each site. Electrons in 4f doublets are strongly localized and do not form a band; its single or double occupancy corresponds to 4f$^5$ and 4f$^6$ configurations. The 5d doublet, on the other hand hybridize with nearest neighbors and forms a band, very similar to 5d bands in LaB$_6$ and CeB$_6$.

Total spin is no more a good quantum number. However, strong quantum fluctuations of the \emph{Kramer spins} results in a \emph{paramagnetic state}. So as a starting point we consider the SU(2) spin invariant Kondo lattice model Hamiltonian.

Majorana fermion representation has been valuable in theories of two channel Kondo effect \cite{EmeryKivelson}, Luttinger liquids \cite{Affleck}, Kondo insulators, quantum spin liquids \cite{spinliquid} and Kitaev models \cite{KitaevHoneyComb}. A spinful free fermi sea can be rewritten exactly \cite{ColemanMirandaTsvelik} interms of one spinless scalar Majorana fermi sea and three spinful Majorana fermi sea.  Consider a tight binding model of electrons, with a nearest neighbor hopping, in a bipartitle lattice at half filling. Transform operators in one of the two sublattices,  $c^\dagger_{i\sigma} \rightarrow i c^\dagger_{i\sigma}$ to get:
\be
H_{KE} = - t \sum_{\langle ij \rangle} (c^\dagger_{i\sigma} c_{j\sigma} + c^\dagger_{j\sigma} c_{i\sigma}) \rightarrow
- i t \sum_{\langle ij \rangle} (c^\dagger_{i\sigma} c_{j\sigma} - c^\dagger_{j\sigma} c_{i\sigma})
\ee
Using four Majorana fermions, a scalar $c_{0i}$ and a vector ${\bf c}_{i} \equiv ( c_{ix}, c_{iy}, c_{iz}$),  we rewrite electron operators as: $ c^\dagger_{i\uparrow} = \tfrac{1}{{2}} (c_{ix} + i c_{iy})$ and $ c^\dagger_{i\downarrow} = \tfrac{1}{{ 2}} (c_{iz} - i c_{i0})$. From the anticommutation property of electron operators, the anticommutation properties of Majorana fermions follow:
$ \{c_{i\alpha},c_{j\beta}\} = 0$ for $ i \neq j$ and $ \alpha \neq \beta$ and $c_{i\alpha}^2 = 1$. In terms of the scalar and vector Majorana fermions kinetic energy takes an elegent SU(2) spin rotational invariant form form:
\be
H_{KE} = - i t \sum_{\langle ij \rangle} (c_{i0}c_{j0} + {\bf c}_{i} \cdot {\bf c}_{j})
\ee
To diagonalize the above interms of positive energy physical (complex fermion) excitations, Majorana fermions $c_{i\alpha}$ at site ${\bf R_i}$ can be written interms of (canonical) complex fermions $(a^\dagger_{\bf k \alpha}, a^{}_{\bf k \alpha})$, $\alpha, = 0,x,y,z$ , defined in half of the Brilluouin zone;
$c_{i\alpha} = \frac{1}{\sqrt N} \sum_{\rm half~BZ} (a_{\bf k \alpha} e^{i{\bf k \cdot R}_i} + a^\dagger_{\bf k \alpha}e^{-i{\bf k \cdot R}_i})$. Using this relation we get the kinetic energy Hamiltonian of a half filled band of electrons:
\be
 H_{KE} = \sum_{k \in {\rm half~BZ}}\epsilon_k  (a^\dagger_{k0}a_{k0} + {\bf a}^\dagger_{k} \cdot {\bf a}_{k}) + E_{0}
\ee
This is an exact representation of free electrons interms of complex fermions with positive excitation energies $\epsilon_k$, energy dispersion in a cubic lattice. These 4 species of physical fermions are their own antiparticles and are defined in the outside of the reference fermi surface in k-space. Negative energy states are absent.

The constant E$_{0} = 2 \sum_{\epsilon_k \leq 0} \epsilon_k = 2N \int_{-W}^{0} \epsilon \rho(\epsilon) d\epsilon $ is the ground state energy of the half filled band of physical electrons. Here 2W is the band width and ${\rho (\epsilon)}$ is the one electron density of states. It is easy to see from their construction, or by performing a global U(1) transformation that the new complex fermions do not have sharp and definite charge. They are similar to neutral Bogoliubov quasi particles. Electric and magnetic fields couple to these excitations in a counter intuitive fashion, as we will see later.

We also note that the scalar Majorana fermion is constructed from down spin electrons: $c_{i0} = c^{\dagger}_{i\downarrow} + c^{}_{i\downarrow}$. Choice of down spin is arbitrary and final result is independent of choice of spin direction.  Further, Majorana fermion kinetic energy, when expressed in terms of electrons contains BCS type of pairing (anomalous) terms, such as $c^{\dagger}_{i\downarrow} c^{\dagger}_{j\downarrow}$. However, coefficient of this term t is fixed and there is no phase degree of freedom.

Majorana fermion decomposition of the standard spinful electron fermi sea at half filling is a way of organizing quantum fluctuations and excitations interms of new degrees of freedom. In particular, the scalar Majorana fermion, which will play important role for us, selects out fluctuating charge component of electrons, and separates it from spin fluctuation degree of freedom.

The SU(2) symmetric Kondo lattice Hamiltonian is :
\be
H_{} = H_{\rm KE} + J \sum_{i} {\bf s}_i \cdot {\bf S}_i
\ee
where ${\bf S}_i$ is the operator of the spin-half local moment. It can be written \cite{ColemanMirandaTsvelik} interms of a vector Majorana fermion ${\boldsymbol \eta}_i$ as $ {\boldsymbol S}_i \equiv -\frac{i}{2} {\boldsymbol \eta}_i\times{\boldsymbol \eta}_i$. In this representation there are no local constraints. However, there is a Z$_2$ gauge redundancy and  dimension of the local moment Hilbert space gets enlarged by a factor $2^{\frac{N}{2}}$, where N is the number of sites.

Conduction electron spin operator at the i-th site is,  ${\bf s}_{i} \equiv  c^{\dagger}_{i\alpha} {\boldsymbol \sigma}_{\alpha \beta} c^{}_{i\beta}$. Interms of Majorana fermions it becomes a sum of two vectors:
\be
{\bf s}_i = i c_{i0} {\bf c}_{i} + \tfrac{i}{2}  ({\bf c}_{i} \times {\bf c}_{i})
\ee
Further, using the identity, $({\bf c}_{i} \times {\bf c}_{i}) \cdot ({\boldsymbol \eta}_i\times{\boldsymbol \eta}_i) = - \tfrac{1}{2} ({\bf c}_i \cdot {\boldsymbol \eta}_i)^2 - \tfrac{3}{2} $, we get
\be
H_{} = H_{\rm KE} +
{\tfrac{J}{2}} \sum_i[c_{i0} {\bf c}_{i} \cdot ({\boldsymbol \eta}_i \times {\boldsymbol \eta}_i) - \frac{1}{2}({\bf c}_{i} \cdot {\boldsymbol \eta}_i)^2] + {\rm const~}
\ee
The above representation of Kondo lattice Hamiltonian is exact, except for the gauge redundancy. That is, many body spectrum of the original Kondo Hamiltonian is preserved. However there are 2$^{\frac{N}{2}}$ extra gauge copies of the Hilbert space.

Further, we have rewritten the Kondo lattice Hamilonian in a form different from CMT \cite{ColemanMirandaTsvelik}. Our rewiring help us see emergence of a free scalar Majorana fermion in a straight forward Hartree type factorization, without a need to invoke odd frequency pairing ansatz explicitly.

In this paper we focuss on non magnetic solution with vanishing vector order parameters: $\langle c_{i0} {\bf c}_{i}\rangle$ = 0 and $\langle {\bf S}_i \rangle = \langle {\boldsymbol \eta}_i \times {\boldsymbol \eta}_i\rangle = 0$. In this approximation the first term in equation 5 drops out. Second term becomes $({\bf c}_{i} \cdot {\boldsymbol \eta}_i)^2 \rightarrow 2 \chi_0 ({\bf c}_{i} \cdot {\boldsymbol \eta}_i) - \chi_0^2$. Here $\chi_0 \equiv \langle {\bf c}_{i} \cdot {\boldsymbol \eta}_i\rangle$,   Mean field Hamiltonian and its diagonalized form become:
\bearr
H_{\rm mf}& =& H_{\rm KE} +
- J \chi_0 \sum_i {\bf c}_{i} \cdot {\bf \eta}_i+ {\rm const~} \nonumber \\
&=& \sum_{\rm Half~BZ} \epsilon_k a^\dagger_{k0} a^{}_{k0} + \sum_{BZ} {\varepsilon}_k {\bf \tilde a}^{\dagger}_k.{\bf \tilde a}^{}_k + {\rm const~}
\eearr
Scalar Majorana fermions remain unaffected. The new vector (complex) fermion operators (${\bf \tilde a}^{\dagger}_{k}, {\bf \tilde a}^{}_{k}$) are hybrids of vector Majorana fermions $ {\bf c}_i$'s of conduction electrons and Majorana fermions ${\boldsymbol \eta}_i$'s  of local moments. The hybrid vector (complex) fermion is defined in the entire BZ, as local moments have become dynamical fermions leading to doubling of vector fermion degree of freedom.  Interaction modified energy dispersion of the (complex) vector fermions is, $\varepsilon_{k} = \frac{\epsilon_k}{2} \pm \sqrt { (\frac{\epsilon_k}{2})^2 + (J\chi_0)^2}$, exhibiting a hybridization gap.

The above mean field ansatz breaks a local Z$_2 \times$ Z$_2$ symmetry, enjoyed by the conduction electron Majorana fermions and local moment Majorana fermions, down to a local Z$_2$ symmetry. Vector Majorana spin of the conduction electron at a given site ${\bf c}_i$ and vector Majorana fermion of local moment ${\boldsymbol \eta}_i$ get locked as parallel vectors and quantum fluctuate. As for as spin part is concerned we have a gapped spin liquid state that supports gapful spin-1 neutral fermionic excitations.

We find that there are non topological localized soliton excitations in which i) the condensate $\chi_0 ({\bf R}_i)$ varies in space (for example vanishing at a given site) and selfconsistently liberates a Kondo spin from exchange coupling with conduction electrons and ii) $\langle c_0i {\bf c}_i \rangle $ and or  $\langle {\boldsymbol \eta}_i \times {\boldsymbol \eta}_i \rangle $ get non zero expectation value (for example non zero at a given site) and traps zero energy vector Majorana modes. We will not go into this in this paper.

One of our main objectives is to show that scalar Majorana fermi liquid that survives Kondo interaction in \smb responds to uniform magnetic field very much like a standard fermi liquid. In other words, our \emph{scalar neutral Majoran fermi liquid exhibits Landau diamagnetism}. 

Torque magnetometry dHvA experiments (and quantities such as specific heat etc.) measure how internal energy U(H) of an electron system changes with magnetic field.  In particular magnetic moment $ M = - \frac{\partial U}{\partial H}$, where U is the internal energy. These and sound attenuation measurements are sensitive to variation of density of states of excitations at the fermi level.  Shubnikov dHvA oscillations in resistivity is a different story.

In what follows we calculate internal energy of the fermi sea, first without Kondo interaction. This simple calculation is already revealing.
Eventhough fermionic excitations of the four Majorana fermi seas (one scalar and three vector) are nominally neutral each one contributes
equally to magnetic field dependence of internal energy, as if they are charged ! The sum of contributions from the 4 nominally neutral Majorana fermi seas addup to the value we obtain for the standard charged feermi sea of electrons with spin degree of freedom.

To see this we consider the kinetic energy term, in the presence of uniform magnetic field ${\bf H} = {\boldsymbol \nabla} \times {\bf A}$.  We diagonalize the kinetic energy term interms of one particle Landau orbitals $\alpha$ and their eigen values $\epsilon_\alpha$ of the tight binding lattice:
\bearr
H_{\rm KE}& =& - t \sum_{\langle ij \rangle} (e^{i\frac{e}{\hbar c}{ \int_i^j {\bf A}\cdot d{\bf l}}} c^\dagger_{i\sigma} c^{}_{i\sigma} + H.c.) \nonumber \\
&=& \sum_{{\rm all}~\alpha} \epsilon_{\alpha} c^{\dagger}_{\alpha \sigma}c^{}_{\alpha \sigma}
\eearr
Energy of the half filled band of (up and down spin) fermi sea gets modified from $ 2\sum_{\epsilon_k \leq 0} \epsilon_k$ to $ 2\sum_{\epsilon_{\alpha} \leq 0} \epsilon_\alpha$, in the presence of a finite uniform magnetic field. Magnetic field preserves the particle hole symmetry.  We first rewrite the electron operators (equation 8) in terms of Majorana fermions  $(c_{0\alpha}, {\bf c}_\alpha )$ expressed in the Landau orbital basis and then diagonalize the Hamiltonian to write it interms of positive energy complex scalar Fermions, $(a^{\dagger}_{\alpha}, a^{}_{\alpha})$ and vector fermions $({\bf a}^{\dagger}_{\alpha}, {\bf a}^{}_{\alpha})$ (similar to equation 8) and obtain
\bearr
H_{\rm KE}=   \sum_{\epsilon_\alpha \geq 0 }\epsilon_{\alpha} {a}^{\dagger}_{\alpha 0} {a}_{\alpha 0}  + \sum_{\epsilon_\alpha \geq 0 }\epsilon_{\alpha} {\bf {a}}^\dagger_{\alpha} \cdot {\bf {a}}^{}_{\alpha} + E_0({\bf H})
\eearr
Vacuum energy of individual Majorana fermi seas are all identical. Total vacuum energy is, E$_0 ({\bf H}) = 4 \times \tfrac{1}{2} \sum_{\rm \epsilon_\alpha \leq 0 }\epsilon_{\alpha} \equiv
2N \int^0_{-W} \rho (\epsilon, {\bf H}) \epsilon d\epsilon$. Here $\rho (\epsilon, {\bf H})$ is the modified one electron density of states in the presence of an uniform magnetic field $\bf H$ and 2W is the bandwidth of one electron states.

As our fermi surfaces are non-spherical, vacuum energy E(${\bf H}$) changes as the strength and direction of the  magnetic field is varied, leading to dHvA oscillations, through change in thermodynamic quantities. Our simple and new result is that \emph{all Majorana fermions including the vector and scalar ones, though apparently charge neutral, contribute equally to quantum oscillations (Landau Diamagnetism) in the non interacting case}. Physically it follows from that fact our Majorana fermions are quantum coherent combinations of charge carrying electrons and holes.

Armed with this insight we study the interacting case in the presence of magnetic field. We present our mean field theory briefly. Since Kondo interaction is an on site interaction and hybridization gap $J\chi_0 << \hbar \omega_c$, for experiments in \smb,  we use same translationally invariant meanfield ansatz as before: $\langle {c_{i 0}{\bf c}_{i}\rangle}= 0,~\langle {\boldsymbol \eta}_i \times {\boldsymbol \eta}_i \rangle = 0$  and $\langle {\bf c}_{i} \cdot {\boldsymbol \eta}_i \rangle  = \chi_0$. We also ignore the small zeeman splitting term. The new mean field Hamiltonian is:
\bearr
H_{\rm mf}({\bf H}) &=& \sum_{\epsilon_\alpha \geq 0} \epsilon_\alpha a^\dagger_{\alpha 0} a^{}_{\alpha 0} + \sum_{\rm \epsilon_\alpha \geq 0 }\epsilon_{\alpha} {\bf a}^\dagger_{\alpha} \cdot {\bf a}^{}_{\alpha} - J \chi_0 \sum_i {\bf c}_{i} \cdot {\boldsymbol \eta}_i \nonumber \\
&+& \tfrac{1}{4} E_0({\bf H}) + \tfrac{3}{4} E_v ({\bf H}, \chi_0)
\eearr
Here $\frac{1}{4} E_{0} ({\bf H})$ is the vacuum energy of the scalar Majorana fermi sea, which remains unaffected by Kondo interation. And $\frac{3}{4} E_{v} ({\bf H}, \chi_0)$ is the interaction and magnetic field modified vacuum energy of the 3 vector Majorana fermions. As the mean field interaction term is a simple sum over sites, the same Landau basis diagonalizes the mean field Hamiltonian leading to the final form:
\bearr
H_{\rm mf}({\bf H}) &=& \sum_{\epsilon_\alpha > 0} \epsilon_\alpha a^\dagger_{\alpha 0} a^{}_{\alpha 0} + \sum_{\rm all~{\varepsilon}_\alpha }
{{\varepsilon}_{\alpha}} {\bf {\tilde a}}^\dagger_{\alpha} \cdot {\bf {\tilde a}}^{}_{\alpha} \nonumber \\
&+& \tfrac{1}{4} E_0({\bf H}) + \tfrac{3}{4} E_v ({\bf H}, \chi_0)
\eearr
Energy spectrum of the interaction modified hybridized (complex) vector fermions $({\bf {\tilde a}}^\dagger_{\alpha}, {\bf {\tilde a}}^{}_{\alpha})$ becomes  ${{\varepsilon}}_{\alpha} = \frac{\epsilon_\alpha}{2} \pm \sqrt { (\frac{\epsilon_\alpha}{2})^2 + (J\chi_0)^2}$. We do not discuss the gap equation for the order parameter $\chi_0$ in the present paper and compare mean field energies with other mean field solutions.

As the hybridization energy given by the scale J$\chi_0 >> \hbar \omega_c$, optimal value of $\chi_0$ has a very weak dependence on the applied magnetic field. Consequently vacuum energy of the vector Majorana fermions have very weak dependence on the magnetic filed. However, the scalar fermions continue to contribute $\frac{1}{4}$ of a free electron gas value at the same magnetic field.  This is our explanation for the observed quantum oscillations in the bulk of \smb. To explain some missing bands etc., in comparison to LaB$_6$ and SmB$_6$, one needs to go into realistic models.

We first offer a physical explanation for the survival of scalar Majorana fermi sea in the presence of Kondo interaction. We call this an \emph{emergent 2-channel Kondo phenomenon in a one channel system} \cite{CIT}. When Kondo coupling is small, J $<<$  2W (electron band width), fermi surface gets modified around a small energy shell in k-space of energy width $ \sim J$.  Vector Majorana fermions and scalar ones in this shell are actively involved in interaction with Kondo spins. However interaction affects them in different fashion.

At any given site, free fermi gas at half filling has equal probability $\frac{1}{4}$ for occurrence of neutral spin states $|\uparrow\rangle, |\downarrow \rangle$ and charged pseudo spin states $|\uparrow\downarrow\rangle, |0\rangle$.  Physical spin operator ${\bf s}_i$ acts on neutral spin states. The pseudo spin
${\boldsymbol \tau}_i \equiv [ (c^\dagger_\uparrow c^\dagger_\downarrow + c^{}_\downarrow c^{}_{\uparrow}), i (c^\dagger_\uparrow c^\dagger_\downarrow - c^{}_\downarrow c^{}_\uparrow), \frac{1}{2} (c^\dagger_{i\uparrow} c^{}_{i\uparrow} + c^\dagger_{i\downarrow} c^{}_{i\downarrow} - 1)]$ acts on the doubly occupied and empty states. Vector and scalar Majorana fermions organize fluctuations among these 4 configurations in a fashion that helps us to solve our problem.

Spinful, singly occupied neutral conduction electron states at a given site alone couples with local moment at that site. So we naively expect enhanced singlet correlations (quantum entanglement) between electron spins and Kondo spins in the many body ground state wave function; and absence of correlation (quantum entanglement) between pseudo spin and the Kondo spin. Such a wave function in configuration space will have, in general, sharp variations leading to increased electron kinetic energy and Kondo spin fluctuation energies. One way to reduce this sharp variations is to utilise the particle hole (spin-pseudo spin, SU(2)$\times$SU(2)) symmetry in the problem and entangle Kondo spins and pseudo spins, as if a coupling between Kondo spin and pseudo spin also exists with the same strength J.

In other words, if we manage to generate a symmetric 2 channel effective interaction so that conduction electron spin and pseudo spin couples to Kondo spin with the same strenth, we may be able to construct a more smooth (better) manybody state and gain energy. This indeed happens in our problem, once we recognize that second term in the Kondo interaction in equation 6 (that also survives mean field factorization), rewritten interms of conduction electron variables is $ \tfrac{J}{2} \sum_{\langle ij \rangle} i ({\bf c}_{i}\times{\bf c}_{j})\cdot {\bf S}_i = \tfrac{J}{2} \sum_{\langle ij \rangle} ({\bf s}_i + {\boldsymbol\tau}_{i} )\cdot {\bf S}_i$.

It means that the many body wave function develops a quantum entanglement characteristic of symmetric 2 channel Kondo effect and gains energy. \emph{The pseudo spin acts like a ghost channel}. In this approximation scalar Majorana fermion $c_{0i}$, at every site falls out of the picture. Further hopping of the scalar Majorana fermions $c_{i0}$'s leads to a scalar Majorana fermi sea.

Our proposal of presence of a scalar Majorana fermi sea in \smb at low temperatures has interesting consequences. In addition to the observed quantum oscillation what are possible signatures of the scalar Majorana fermi sea ? Are there some already available ?

An interesting low temperature saturation of bulk resistivity in \smb seen in experiments has been suggested \cite{ResistivitySaturation} to be an artefact of surface conduction phenomenon. Further in the quantum oscillation experiment \cite{SuchitraSmB6QOsc} a strong deviation from Lifshitz Kosevich formula is seen in the temperature regime. \emph{We suggest that the deviation from Lifshitz Kosevich formula and resistivity saturation are  related}. More importantly we suggest an alternate possibility for resistivity saturation, which is intrinsic to the bulk. A scalar Majorana fermion, even though neutral in an average sense,  has a built in coherent and soft charge fluctuation character. Consequently there seems to be an unitarity limit to scattering of physical electron by Kondo spins, leading to the observed resistivity saturation. Physically, an electron injected close to fermi level looses its spin at a short time scale ($\sim \frac{\hbar}{J}$). However it continues to have charge fluctuations at long time scales, in a quantum coherent fashion. Our preliminary theoretical analysis points to an interesting resistivity saturation \cite{GBUnpublished}, arising the physics of coherent charge oscillations in scalar Majorana fermions.

Same coherent charge fluctuations in a scalar Majorana fermion is likely to leave its signature in, STM, NQR, ARPES, optics, XRay inelastic scattering and others measurements, which are sensitive to fluctuating charges. It will be interesting to provide some theoretical guidance to encourage new experiments. Many know fermi surface anomalies, such as Kohn anomaly in the phonon spectrum, possible nesting instabilities, thermal transport, collective modes etc. also needs to be looked into.

There has been earlier variational approaches to Anderson lattice and Kondo insulator problems, pioneered by Stevens and others \cite{Kikoin}. They are non-trivial and physically motivated wave functions, inspired by Varma-Yafet type of variational wave functions \cite{CMVarma} for Anderson impurity model. Do they contain physics of Majorana fermi liquids ?

It has been suggested \cite{ColemanTKI}that the spin orbit coupling and symmetries make \smb a topological Kondo insulator (TKI). There are already interesting experimental manifestations of the TKI behaviour. Many new surface measurements such as STM tunneling \cite{HoffmanSTMKondoGap}, surface state quantum oscillation \cite{SurfaceQOscillation}, ARPES \cite{ARPES}, transport etc. needs to be relooked at, from the point of view of our proposal. 

Our preliminary study shows \cite{GBUnpublished} that there are more exotic possibilities, including soft scalar and vector Majorana fermion excitations at the surfaces. A Fano (antisymmetry) line shape (anti resonance) has been observed \cite{HoffmanSTMKondoGap}. This is likely to be connected to the underlying Majorana fermion character of low energy quasi particles.

In the present article we have focussed on how a scalar Majorana fermi liquid emerges and did not discuss possible p-wave odd frequency pairing aspect. This needs to be looked into, as there are interesting predictions by CMT \cite{ColemanMirandaTsvelik}.

Our proposal calls for a relook at well known Kondo insulators, including SmS and FeSi. It is likely that some of the known Kondo insulators might support gapless Majorana fermi liquids. Nature of doped Kondo insulator or doped Majorana Fermi liquid is an intereseting quation. It will be interesting to see if scalar and Majorana fermi liquid excitations continue to survive after doping. 

A light mass metallic surface band in \smb is seen in the experiments \cite{SurfaceQOscillation,ARPES}. There is a discrepancy between theory and experiments \cite{ColemanSurfaceBand}. Theory shows a heavy mass.  We believe that there is a deep connection of the observed surface band to the bulk scalar Majorana fermi sea we have proposed. Lines shaps in STM performed on \smb has Fano line shape. 

To conclude, we have proposed that the observed quantum oscillations in \smb arise from a scalar Majorana fermi liquid. It is a novel quantum state that was indeed suggested in an earlier mean field theory of Kondo insulator by CMT \cite{ColemanMirandaTsvelik}. We have reformulated the mean field theory in a simple fashion to make the appearance of a scalar Majorana fermi sea natural, without explicitly involving odd frequency pairing. Most importantly we show that the scalar Majorana fermi sea, even though apparently charge neutral, couples to external magnetic field exactly in the same fashion that a charged fermi liquid does. We have discussed the physics behind an emergent compactified 2-channel Kondo interaction that underlies physics of a scalar Majorana fermi sea formation and gap for spin-1 and charged excitations.

 If our explanation of the observed quantum oscillations in insulating \smb is proved correct by further experiments and theory, it is likely to bring in new physics to the exciting field of Kondo insulators and heavy fermion physics and strongly correlated electron systems.

\emph{Acknowledgement}: It is a pleasure to thank Suchitra Sebastian for bringing to my attention their exciting quantum oscillation results before publication. I thank her and Mukul Laad for discussions. Years ago I had a coherent exposure to correlated insulators and valence fluctuations from A. Jayaraman, C.M. Varma, R. Nityananda, N. Kumar and H.R. Krishnamurthy at Bangalore. I thank them. I thank Science and Engineering Research Board (SERB, India) for a National Fellowship. This work is supported by the Government of Canada through Industry Canada and by the Province of Ontario through the Ministry of Research and Innovation.

\end{document}